\documentstyle[12pt,epsf,epsfig]{article}
\newcount\timecount
\newcount\hours \newcount\minutes  \newcount\temp \newcount\pmhours

\hours = \time
\divide\hours by 60
\temp = \hours
\multiply\temp by 60
\minutes = \time
\advance\minutes by -\temp
\def\hour{\the\hours}
\def\minute{\ifnum\minutes<10 0\the\minutes
            \else\the\minutes\fi}
\def\clock{
\ifnum\hours=0 12:\minute\ AM
\else\ifnum\hours<12 \hour:\minute\ AM
      \else\ifnum\hours=12 12:\minute\ PM
            \else\ifnum\hours>12
                 \pmhours=\hours
                 \advance\pmhours by -12
                 \the\pmhours:\minute\ PM
                 \fi
            \fi
      \fi
\fi
}

\def\monthname{\relax\ifcase\month 0/\or January\or February\or
   March\or April\or May\or June\or July\or August\or September\or
   October\or November\or December\else\number\month/\fi}

\def\bold#1{\setbox0=\hbox{$#1$}%
     \kern-.025em\copy0\kern-\wd0
     \kern.05em\copy0\kern-\wd0
     \kern-.025em\raise.0433em\box0 }

\def\gappeq{\mathrel{\rlap {\raise.5ex\hbox{$>$}}
{\lower.5ex\hbox{$\sim$}}}}

\def\lappeq{\mathrel{\rlap{\raise.5ex\hbox{$<$}}
{\lower.5ex\hbox{$\sim$}}}}

\def\ga{\mathrel{\raise.3ex\hbox{$>$\kern-.75em\lower1ex\hbox{$\sim$}}}}
\def\la{\mathrel{\raise.3ex\hbox{$<$\kern-.75em\lower1ex\hbox{$\sim$}}}}
\def\gev{{\rm \, Ge\kern-0.125em V}}
\def\tev{{\rm \, Te\kern-0.125em V}}
\def\beq{\begin{equation}}
\def\eeq{\end{equation}}

\def\ohsq{\Omega_{\chi} h^2}

\def\m12{m_{1\!/2}}

\begin{document}
\begin{titlepage}
\pagestyle{empty}
\baselineskip=21pt
\rightline{hep-ph/0102331}
\rightline{CERN--TH/2001-054}
\rightline{ACT-02/01, CTP-TAMU-06/01}
\rightline{UMN--TH--1943/01, TPI--MINN--01/12}
\vskip 0.05in
\begin{center}
{\large{\bf
Combining the Muon Anomalous Magnetic Moment with other Constraints
on the CMSSM}}
\end{center}
\begin{center}
\vskip 0.05in
{{\bf John Ellis}$^1$, 
{\bf D.V. Nanopoulos}$^2$ and
{\bf Keith A.~Olive}$^{1,3}$
\vskip 0.05in
{\it
$^1${TH Division, CERN, Geneva, Switzerland}\\
$^2${Department of Physics, Texas A \& M University,
College Station, TX~77843, USA; \\
Astroparticle Physics Group, Houston
Advanced Research Center (HARC), \\
Mitchell Campus,
Woodlands, TX~77381, USA; \\
Chair of Theoretical Physics,
Academy of Athens,
Division of Natural Sciences,  
28~Panepistimiou Avenue,
Athens 10679, Greece}\\
$^3${Theoretical Physics Institute, School of Physics and Astronomy,\\
University of Minnesota, Minneapolis, MN 55455, USA}\\
}}
\vskip 0.05in
{\bf Abstract}
\end{center}
\baselineskip=18pt \noindent

We combine the constraint suggested by the recent BNL E821 measurement of
the anomalous magnetic moment of the muon on the parameter space of the
constrained MSSM (CMSSM) with those provided previously by LEP, the
measured rate of $b \to s \gamma$ decay and the cosmological relic density
$\ohsq$. Our treatment of $\ohsq$ includes carefully the direct-channel
Higgs poles in annihilation of pairs of neutralinos $\chi$ and a complete
analysis of $\chi - {\tilde \ell}$ coannihilation. We find excellent
consistency between all the constraints for $\tan \beta \ga 10$ and $\mu >
0$, for restricted ranges of the CMSSM parameters $m_0$ and $m_{1/2}$. All
the preferred CMSSM parameter space is within reach of the LHC, but may
not be accessible to the Tevatron collider, or to a first-generation $e^+
e^-$ linear collider with centre-of-mass energy below 1.2~TeV. 

\vfill
\vskip 0.15in
\leftline{CERN--TH/2001-054}
\leftline{February 2001}
\end{titlepage}
\baselineskip=18pt

The recent BNL E821 measurement~\cite{BNL} of the anomalous magnetic
moment of the muon, $a_\mu \equiv (g_\mu - 2)/2$, may indeed be a
harbinger
of new physics~\cite{CM} beyond the Standard Model: $\delta a_\mu \equiv
a_\mu^{exp} - a_\mu^{SM} = (43 \pm 16) \times 10^{-10}$. The largest error
in the Standard Model prediction is that due to the hadronic
contributions, principally vacuum polarization diagrams, with the most
important uncertainty being that in the low-energy region around the
$\rho$ peak.  The value of these hadronic contributions~\cite{Davieretal}
used in the E821 paper~\cite{BNL} does not include the latest data from
Novosibirsk~\cite{Novo}, Beijing~\cite{BES} and CLEO~\cite{CLEO}, but
these are unlikely~\cite{Davierpc} to change the overall picture: we
recall that the hadronic error $\la 7 \times 10^{-10}$ is much smaller
than the apparent discrepancy and the experimental error.  Advocates of
new physics beyond the Standard Model may therefore be encouraged.
However, we recall that the $Z \to {\bar b} b$ branching ratio was once
thought to show a bigger discrepancy with the Standard Model, and we also
caution that the 2.6 $\sigma$ significance of the muon anomaly is formally
less than the preliminary 2.9 $\sigma$ significance of the LEP Higgs
`signal'~\cite{LEPHiggs}. 

{\it A priori}, the BNL measurement favours new physics at the TeV scale,
and we consider the best motivated candidate to be supersymmetry. Even
before the hierarchy motivation for supersymmetry emerged, the potential
interest of $a_\mu$ was mentioned, and a pilot calculation
performed~\cite{Fayet}. Soon after the realization that supersymmetry
could alleviate the hierarchy problem, the first `modern' calculations of
supersymmetric contributions to $a_\mu$ were
published~\cite{GM,EHN,BM,Kosower}. These were followed by more complete
calculations~\cite{AN,Ven,Abel,LNW} including the mixing expected for
neutralinos, charginos and smuons. In particular, it was noted
in~\cite{LNW} that some contributions are enhanced at large $\tan \beta$. 
The supersymmetric calculations we use in this paper are taken
from~\cite{IN} - for other recent calculations, see~\cite{Moroi}, and we
include the leading two-loop electroweak correction factor~\cite{CM2l}. 
For some time, it has been emphasized~\cite{LNW,CN} that the BNL
experiment would be sensitive to a large range of the parameter space of
the constrained minimal supersymmetric extension of the Standard Model
(CMSSM)  with universal soft superymmetry-breaking parameters at the input
GUT scale, determining in particular the sign of the Higgs mixing
parameter $\mu$~\cite{LNW,CN}. Combining these calculations with the BNL
measurement, $\mu > 0$ is favoured, along with values of $\tan \beta$ that
are not very small. 

The constraints from the E821 experiment are particularly interesting when
combined with the information from LEP~\cite{LEPHiggs}, the measured value
of the $b \to s \gamma$ decay rate~\cite{bsgexpt} and restrictions on cold
dark matter imposed by astrophysics and cosmology, assuming that the
lightest supersymmetric particle (LSP) is the lightest neutralino
$\chi$~\cite{EHNOS}, and that $R$ parity is conserved. Several
combinations of these other constraints have been made by
us~\cite{EFOS,EFOSi,EFGO,EGNO,EFGOSi,LNY} and
others~\cite{others,reviews}, before the advent of the E821 result.

We draw particular attention to a recent combined analysis~\cite{EFGOSi}
of these constraints at large $\tan \beta > 20$, which benefited from
recently available $b \to s \gamma$ calculations~\cite{newbsgcalx} and
made new calculations at large $\tan \beta$ of the the relic density
$\ohsq$.  We found~\cite{EFGOSi} two important effects on the calculation
of $\ohsq$, due to improvements of previous calculations of $\chi -
{\tilde \ell}$ coannihilations and direct-channel $\chi \chi$
annihilations through the heavier neutral MSSM Higgs bosons $H$ and $A$.
Both of these effects extended the region of CMSSM parameter space
consistent with cosmology out to values of the universal soft
supersymmetry-breaking mass parameters $m_0, m_{1/2}$ that were larger
than at $\tan \beta \la 20$. As a result, the discovery of sparticles at
the LHC could not be `guaranteed' in the CMSSM at large $\tan \beta$,
unlike the case when $\tan \beta \le 20$~\cite{EFOSi,EFGO}. Since the recent
BNL measurement favours qualitatively values of $\tan \beta$ that are not
small, as does the LEP Higgs `signal', and since the $b \to s \gamma$
constraint also begins to bite at large $\tan \beta$ even for $\mu > 0$,
it is important to understand the interplay of all these constraints. 

We find good compatibility between all these constraints for $\tan \beta
\ga 10$. Even if one generously allows a 2-$\sigma$ downward fluctuation
in the E821 discrepancy, one finds interesting upper bounds on $m_0$ and
$m_{1/2}$ that effectively extend the previous `guarantee' of CMSSM
discovery at the LHC to large values of $\tan \beta$.  However, no such
`guarantee' can be offered to a linear $e^+ e^-$ collider (LC) with
centre-of-mass energy below 1.2~TeV. We discuss the uncertainties in our
analysis associated with $A_0, m_b$ and $m_t$~\footnote{Several other
papers on the supersymmetric interpretation of the BNL measurement have
already appeared during the last few days, and we comment on them at the
end of this paper.}. 

As already mentioned, our analysis is based on the one-loop calculations
of~\cite{IN}. In relating the masses of the sparticles appearing in the
loops to the basic CMSSM soft supersymmetry-breaking parameters $m_0,
m_{1/2}$, we incorporate the one-loop corrections for charginos and
neutralinos. We also incorporate the leading two-loop electroweak
correction factor $(1 - (4 \alpha / \pi) {\rm ln} ({\tilde
m}/m_\mu))$~\cite{CM2l}, where ${\tilde m}$ is a sparticle mass. We set
the
trilinear soft supersymmetry-breaking parameter $A_0 = 0$ as a default,
but we also discuss the consequences of varying it over the range $- 2
m_{1/2} \le A_0 \le 2 m_{1/2}$. The $b$ and $t$ quark masses enter in our
$m_h$, RGE and relic annihilation calculations. We use as defaults
$m_b(m_b)_{SM}^{\overline {MS}} = 4.25$~GeV and the pole mass $m_t =
175$~GeV, commenting later on the changes as these mass parameters vary
over our allowed ranges $\pm 0.25~\cite{EFGOSi,mb}, \pm 5$~GeV.

In our subsequent discussion, we consider the 2-$\sigma$ range $75 \times
10^{-10} \ge \delta a_\mu \ge 11 \times 10^{-10}$ to be allowed by the
E821 measurement~\cite{BNL}, with the 1-$\sigma$ range $59 \times 10^{-10}
\ge \delta a_\mu \ge 27 \times 10^{-10}$ preferred. We interpret $75
\times 10^{-10}$ as a hard upper limit on $\delta a_\mu$, but models
yielding $\delta a_\mu < 11 \times 10^{-10}$ should perhaps not be
completely excluded yet. We note that a large amount of extra data have
already been taken by E821, and that the present uncertainty will soon be
reduced, which might have dramatic consequences.

The LEP lower limit on the mass of the Higgs boson is $m_h > 113.5$~GeV,
and the possible signal corresponds to $m_h =
115^{+1.3}_{-0.7}$~GeV~\cite{LEPHiggs}. This lower limit applies in the
CMSSM, because the $ZZh$ coupling is unsuppressed relative to the Standard
Model $ZZH$ coupling, unlike in general mixing scenarios possible in the
MSSM. In the following, we display the range $113~{\rm GeV} < m_h <
117~{\rm GeV}$. Given the uncertainties in the Higgs mass
calculations~\cite{HHH},
choices of the MSSM parameters that yield slightly lower values of $m_h$
might be acceptable, whereas values larger than 117~Gev are certainly
allowed if one discards the LEP `signal'.

For $b \to s \gamma$, we allow parameter choices that, after including the
theoretical errors due to the scale and model dependences, may fall within
the 95\% confidence level range $2.33 \times 10^{-4} < {\cal B}(b \to s <
\gamma) < 4.15 \times 10^{-4}$. For the cosmological relic density, we
allow the range $0.1 \le \ohsq \le 0.3$:  the lower bound is optional, as
there may be other sources of dark matter, but the upper bound cannot be
relaxed significantly.

\begin{figure}
\vspace*{-0.75in}
\begin{minipage}{8in}
\epsfig{file=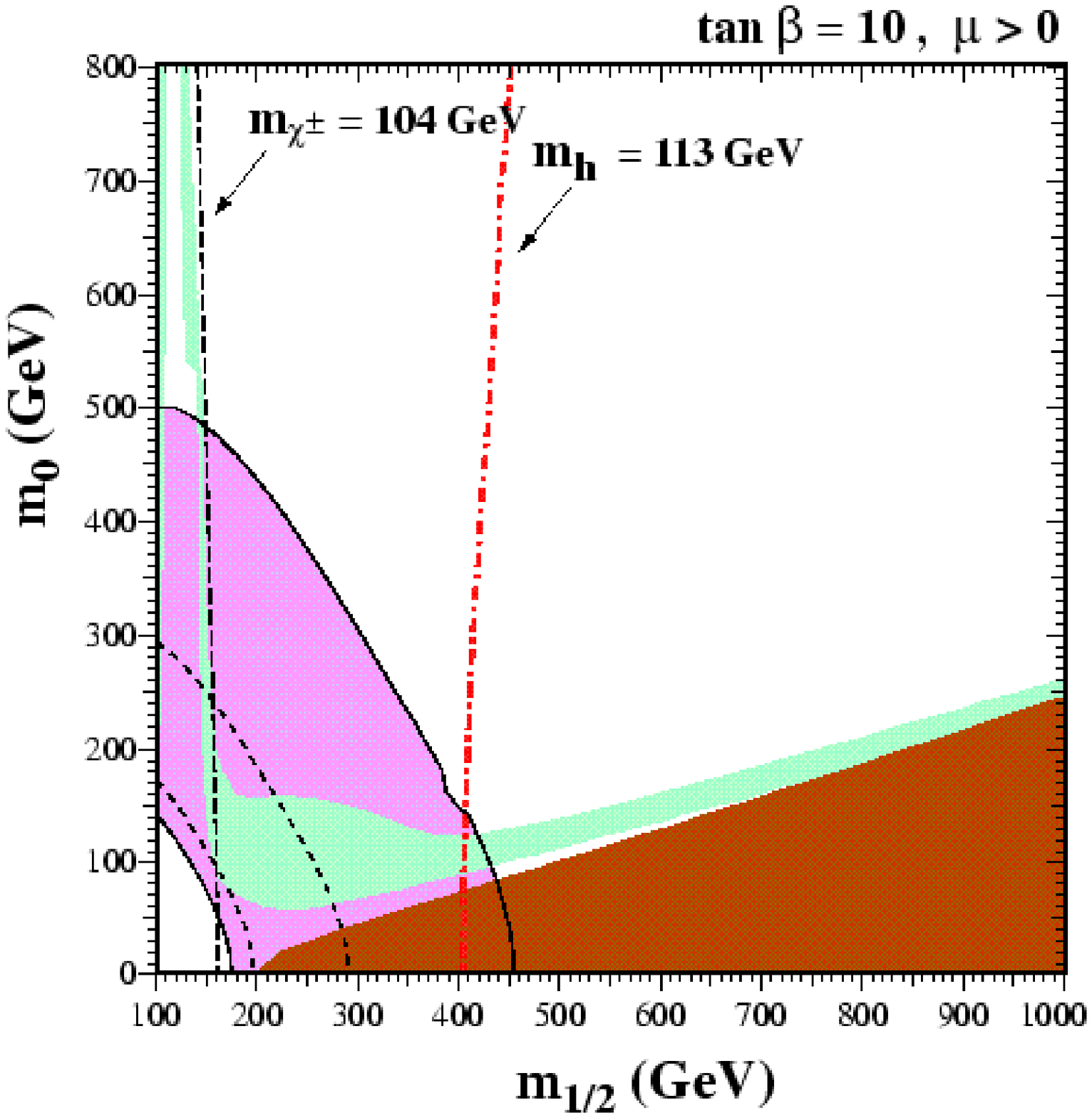,height=3.5in}
\epsfig{file=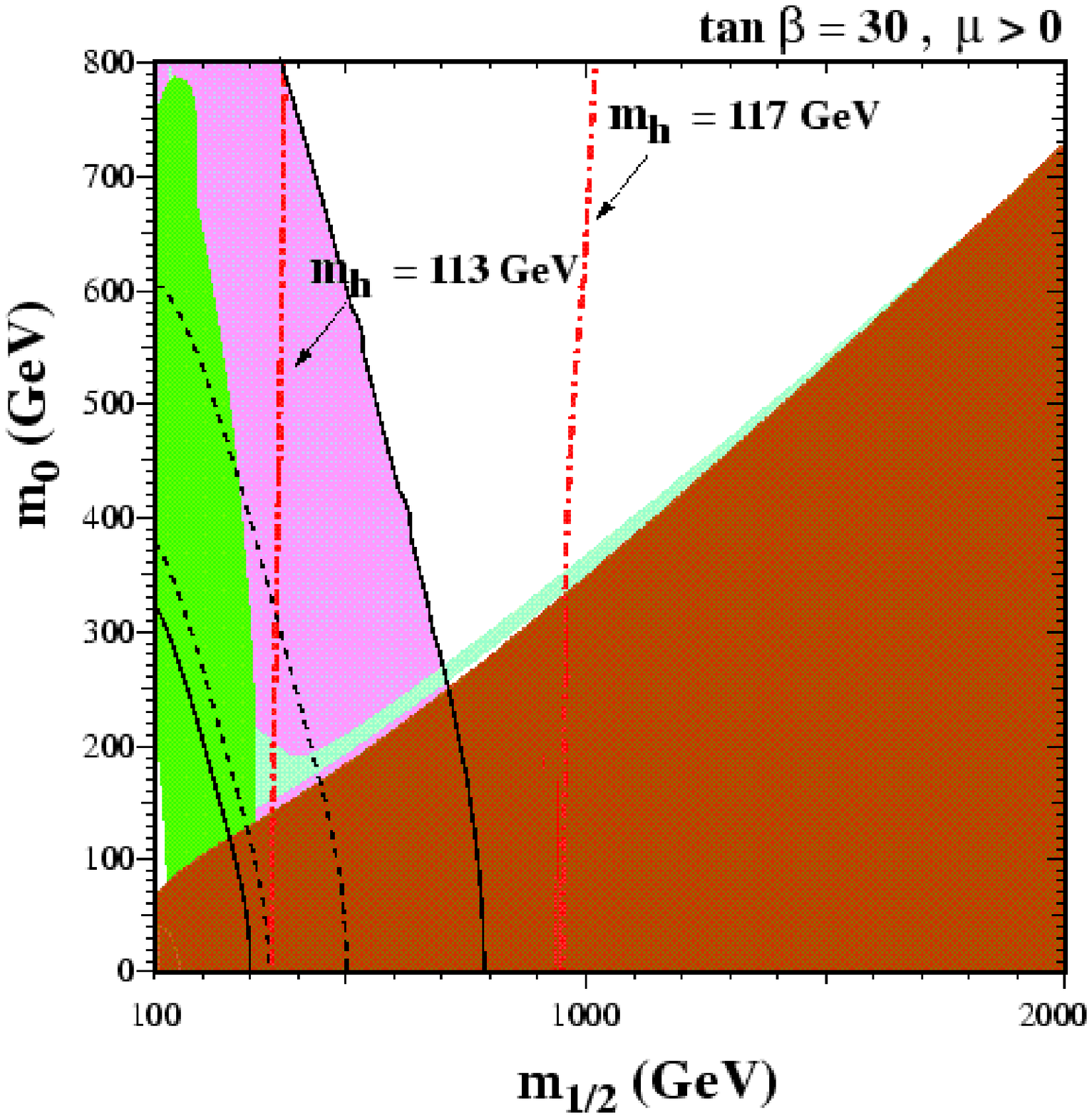,height=3.5in} \hfill
\end{minipage}
\begin{minipage}{8in}
\epsfig{file=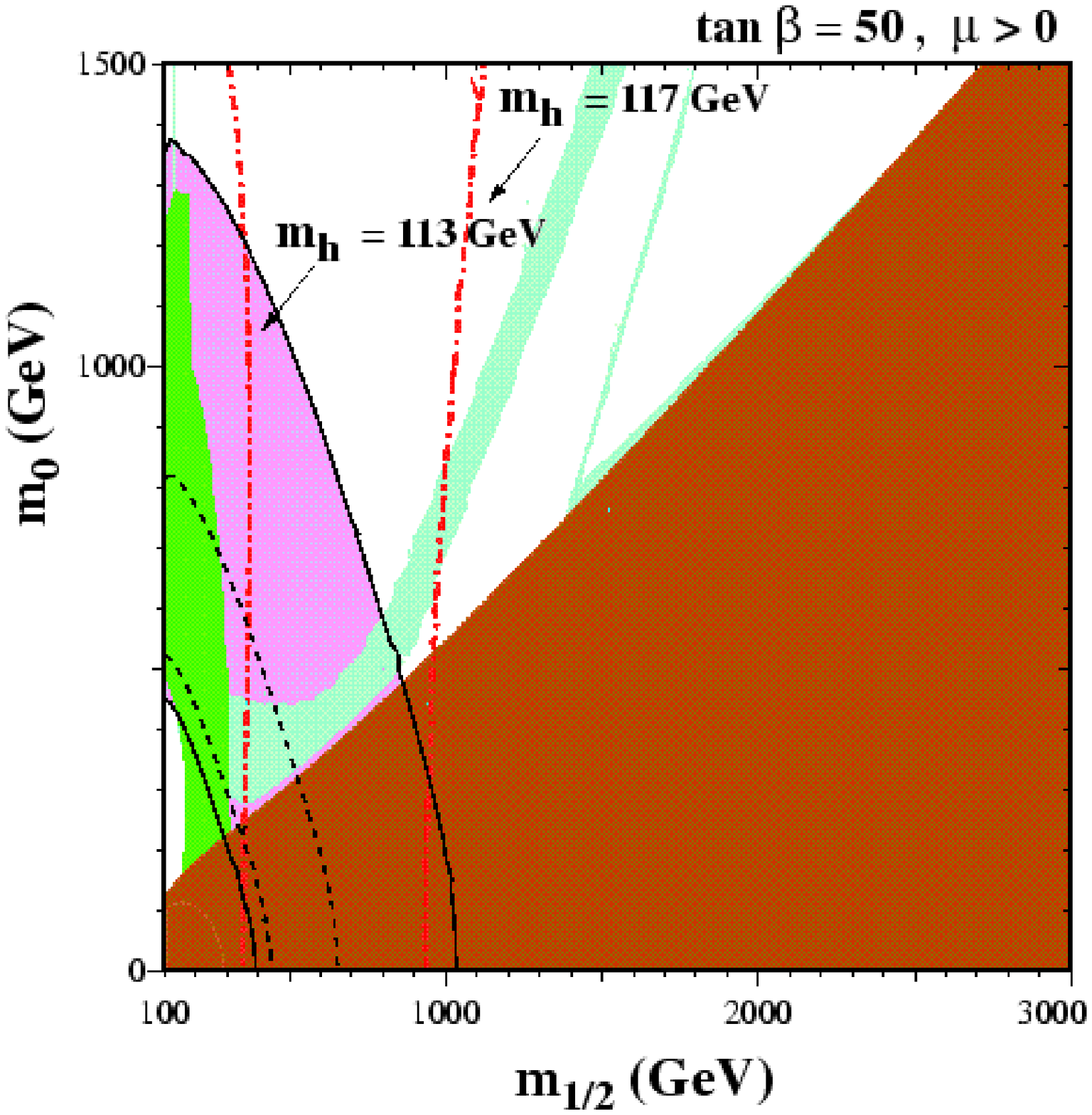,height=3.5in}
\epsfig{file=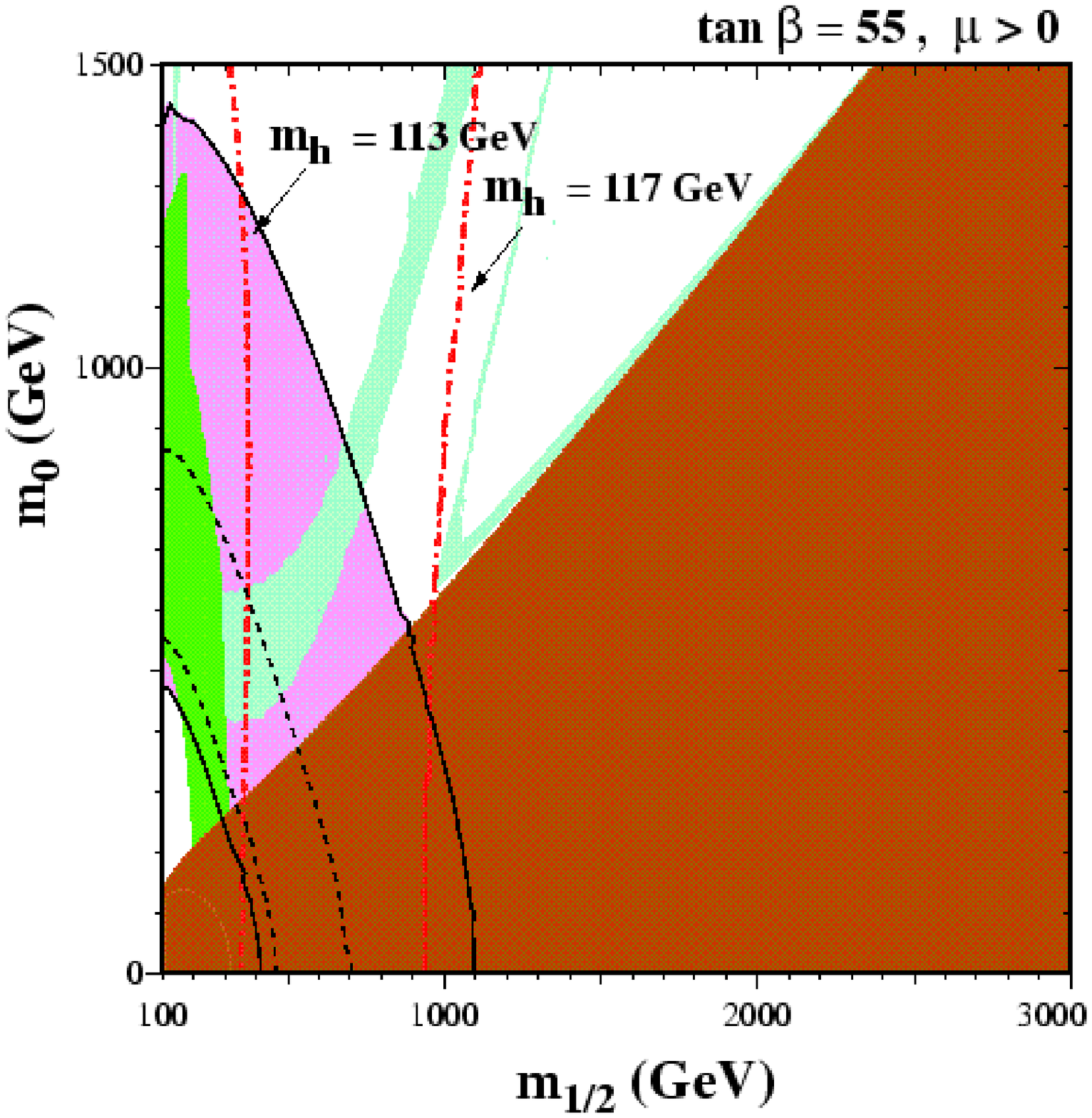,height=3.5in} \hfill
\end{minipage}
\caption{\label{fig:defaults}
{\it The $(m_{1/2}, m_0)$ planes for $\mu > 0$ and $\tan \beta =$ (a) 10,
(b) 30, (c) 50 and (d) 55, found assuming $A_0 = 0, m_t = 175$~GeV and
$m_b(m_b)^{\overline {MS}}_{SM} = 4.25$~GeV. 
The near-vertical (red) dot-dashed lines are the contours
$m_h = 113, 117$~GeV, and the near-vertical (black) dashed line in panel
(a) is the contour $m_{\chi^\pm} = 104$~GeV.
The medium (dark green) shaded regions are excluded by $b
\to s \gamma$.
The light (turquoise) shaded areas are the cosmologically
preferred
regions with \protect\mbox{$0.1\leq\ohsq\leq 0.3$}. In the
dark (brick red) shaded regions, the LSP is the charged ${\tilde \tau}_1$,
so this region is excluded. The regions allowed by the E821 measurement of
$a_\mu$ at the 2-$\sigma$ level are shaded (pink) and bounded by solid
black lines, with dashed lines indicating the 1-$\sigma$ ranges.}}
\end{figure}

We display in Fig.~\ref{fig:defaults} the $(m_0, m_{1/2})$ planes for
representative choices of $\tan \beta$, assuming $\mu > 0$ and $A_0 = 0$.
The regions allowed by the E821 measurement of $a_\mu$ at the 2-$\sigma$
level are (pink) shaded with solid black line boundaries. Also shown as
black dashed lines are the regions favoured by $a_\mu$ at the 1-$\sigma$
level. We display the $m_h = 113, 117~{\rm GeV}$ mass contours as (red)
dash-dotted lines, the dark (green) shaded regions are excluded by $b \to
s \gamma$, the darker (red) shaded regions are excluded because the
lightest supersymmetric particle is the lighter $\tilde \tau$, and the
light (turquoise) regions are where $0.1 \le \ohsq \le 0.3$.  In panel (a) 
for $\tan \beta = 10$, where there is no relevant constraint from $b \to s
\gamma$, we also show as a dashed line the lower limit $m_{\chi^\pm} >
104$~GeV. This excludes the tail of the cosmological region at large $m_0$
and small $m_{1/2}$, where there is rapid annihilation through the $h$
pole. For clarity, this chargino mass contour is not shown in the other
panels, but its effect is similar~\footnote{We do not show the
slepton mass constraint from LEP, which is weaker than the Higgs
constraint in the CMSSM, and is weaker than the upper limit $\delta a_\mu
< 75 \times 10^{-10}$.}.

We observe that there is {\it remarkable} consistency between the
constraints from $a_\mu$, $m_h$, $b \to s \gamma$ and cosmology for $\tan
\beta \ga 10$, as also seen in panels (b, c, d) of
Fig.~\ref{fig:defaults}. Even given the uncertainties in the calculation
of $m_h$, it is difficult to maintain consistency between $a_\mu$ and
$m_h$ for smaller values of $\tan \beta$.  When $\tan \beta \sim 10$, the
other constraints are consistent with cosmology only for $m_0 \sim
100$~GeV, increasing gradually to $m_0 \sim 170$~GeV for $\tan \beta \sim
30$, as seen in panel (b) of Fig.~\ref{fig:defaults}. The favoured range
of $m_0$ increases further as $\tan \beta$ increases, as a result of the
cosmological constraint and the appearance, in particular, of the rapid
$\chi \chi \to A, H$ annihilation process visible in panels (c, d)  for
$\tan \beta = 50, 55$, respectively, which allows $m_0 \la 800$~GeV.  The
allowed ranges of $m_{1/2}$ also increase as $\tan \beta$ increases to
$\sim 50$, where $m_{1/2} \la 900$~GeV is allowed, falling slightly when
$\tan \beta = 55$ because of the rapid $\chi \chi \to A, H$
annihilation~\footnote{Generically, we do not find consistent electroweak
vacua for significantly larger choices of $\tan \beta$, with our default
values of $m_t, m_b$ and $A_0$.}.

The allowed ranges of $m_{1/2}$ are, however, much restricted if one uses
the 1-$\sigma$ range for $a_\mu$, with the maximum value being $\la
500$~GeV. Indeed, combining all constraints and the 1-$\sigma$ range for
$a_\mu$, we find quite small allowed regions of the $(m_{1/2}, m_0)$ plane
centred on:  $\sim (250, 100)$~GeV for $\tan \beta = 10$, $\sim (350,
170)$~GeV for $\tan \beta = 30$, $\sim (400, 350)$~GeV for $\tan \beta =
50$, and $\sim (400, 500)$~GeV for $\tan \beta = 55$~\footnote{Note that,
for $\tan \beta = 10$, we must relax the LEP Higgs constraint to remain
compatible with the 1-$\sigma$ range for $a_\mu$.}.  Typical sparticle
masses
corresponding to these choices are given in the Table.  Comparing with the
CMSSM physics reach for Run II of the Fermilab Tevatron collider, we see
that the trilepton signature may be visible over some fraction of the
allowed region of the $(m_{1/2}, m_0)$ plane for $\tan \beta = 10$, but
not for the larger values of $\tan \beta$ studied here and
in~\cite{Run2sugra}. 

\begin{table}[htbp]
  \begin{center}
    \begin{tabular}{|l|c|c|c|c|c|c|c|c|c|c|c|c|}
\hline
$\tan \beta$&$m_{1/2}$&$m_0$& $m_\chi$& $m_{\chi^\pm}$&$m_{\tilde \tau_1}$
&
$m_{\tilde e_1}$ & $m_{\tilde t_1}$ & $m_{\tilde q}$ & $m_{\tilde g}$&
$m_h$ &
$m_A$&
$\delta a_\mu \times 10^{10}$\\
\hline
10&250&100&99&180&135&145&385&535&580&110&380&30\\
30&350&170&145&270&170& 220& 540& 735& 790& 113& 475& 42\\
50&400&350&170&315&240&385&635&875&895&114&460&40\\
55&400&500&170&315&315&525&665&940&895&114&450& 34\\
\hline
    \end{tabular}
    \caption{\label{tab:tbvmh}
    {\it Typical sparticle masses for points in the $(m_{1/2}, m_0)$ plane
consistent with the 1-$\sigma$ range in $a_\mu$ and other phenomenological
and cosmological constraints. For ${\tilde \tau}$, ${\tilde e}$, and 
${\tilde t}$, the mass corresponds to the lightest eigenstate, which 
is mostly
right-handed, and $m_{\tilde q}$ corresponds to an average slight squark
mass.}}
  \end{center}
\end{table}

As for the LHC, we have shown earlier~\cite{EFOSi} that, in the absence of
the LEP Higgs `signal' and the E821 value of $a_\mu$, cosmology would
allow $m_{1/2} \la 1400$~GeV for $\tan \beta \le 20$ in the coannihilation
region. Studies by members of the CMS Collaboration have shown that at
least some sparticles would be detectable at the LHC throughout this
cosmological region~\cite{CMS}. More recently, however, it
has been shown \cite{EFGOSi} that the maximum value of $m_{1/2}$ increases to
1700 (2200) GeV for $\tan \beta = 30 (50)$, and that the `funnel' of parameters
allowed by rapid $\chi \chi \to A, H$ annihilation also extends out to
large $m_0$ and $m_{1/2}$.  These extensions of the CMSSM parameter space
allowed by cosmology raised the spectre that the LHC might miss
supersymmetry. 

\begin{figure}[htb]
\begin{center}
\mbox{\hskip -.2in \epsfig{file=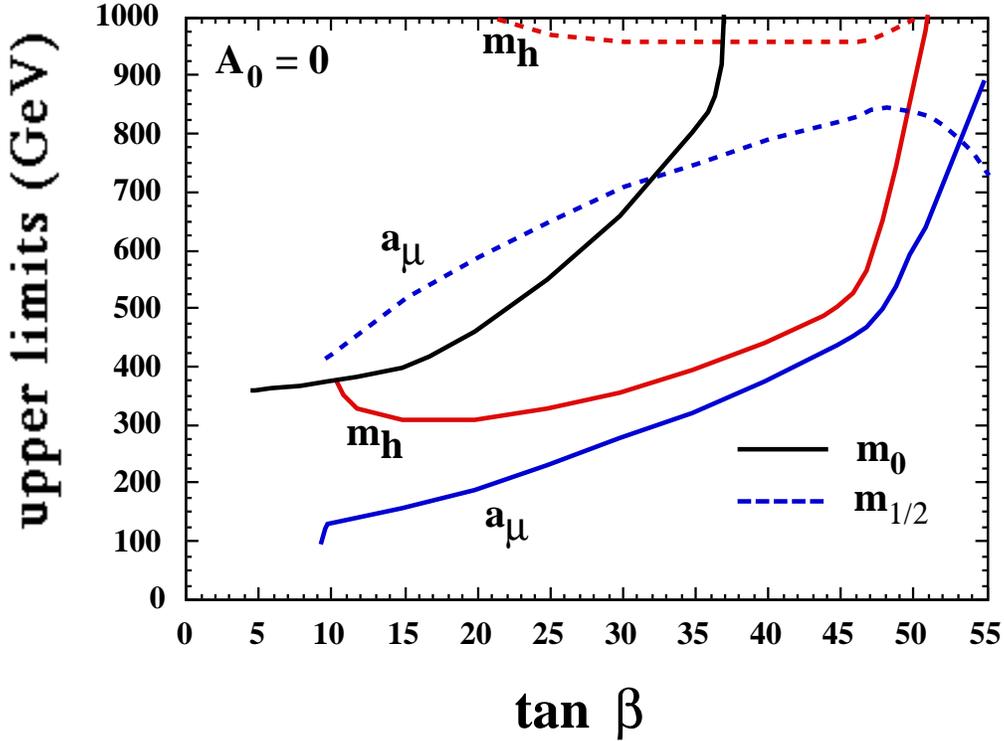,height=10cm}}
\end{center}
\caption[.]{
\label{fig:upper}
{\it Upper limits on $m_{1/2}$ and $m_0$ obtained as functions of
$\tan \beta$ for $\mu > 0$, assuming
$m_b(m_b)^{\overline {MS}}_{SM} = 4.25$, $m_t = 175$~GeV and  $A_0 = 0$.
We show the upper limits on $m_{1/2}$ obtained by combining cosmology
with the LEP Higgs `signal' and the E821 lower limit on $\delta a_\mu$,
and the upper limits on $m_0$ imposed by cosmology alone and in
association with either $a_\mu$ or the LEP Higgs `signal'.}} 
\end{figure}

This is no longer a concern if the E821 lower limit $\delta a_\mu > 11
\times 10^{-10}$ is confirmed. Fig.~\ref{fig:upper} shows the upper limits
on $m_{1/2}$ obtained as functions of $\tan \beta$ by combining cosmology
with E821 or with the upper limit $m_h < 117$~GeV suggested by the
possible LEP Higgs `signal'. The $a_\mu$ constraint is somewhat stronger,
but either would bring supersymmetry back within the range of the LHC. We
also show in Fig.~\ref{fig:upper} the upper limits on $m_0$ imposed by
cosmology alone and in association with the $m_h$ or $a_\mu$ constraints. 
The rapid rise in the upper limit to $m_0$ from cosmology at large $\tan
\beta$ is due to the appearance of the annihilation poles seen in panels
(c) and (d) of Fig.~\ref{fig:defaults}. Below $\tan \beta \simeq 10$, no
independent limit on $m_0$ is provided by the upper bound $m_h < 117 GeV$. 
For $\tan \beta > 10$, the limit on $m_0$ is strengthened gradually as the
$m_h = 117$ GeV contour slides down the coannihilation region, until this
is offset by the shift in the cosmological region to higher $m_0$ as $\tan
\beta$ is further increased.  Eventually, for very large $\tan \beta \ga
45$, the annihilation poles again allow very large values of $m_0$. In
contrast, the lower limit from $a_\mu$ always imposes a significant upper
bound on $m_0$. We conclude that the LHC will find supersymmetry, if the
CMSSM is correct and the E821 lower limit holds up.

It has been commented previously~\cite{EGO} that although discovery of the
CMSSM could be `guaranteed' at the LHC if $\tan \beta \le 20$, there was
no such `guarantee' for a first-generation linear $e^+ e^-$ collider such
as TESLA or the NLC with a centre-of-mass energy below 1.25~TeV, because
of the extension of the cosmologically allowed region by
coannihilation~\footnote{There was also no such `guarantee' in the
focus-point region~\cite{Feng}.}. Since the E821 lower limit $\delta a_\mu
> 11 \times 10^{-10}$ excludes the `tail' of the coannihilation region,
the concern that such a first-generation linear $e^+ e^-$ collider might
miss supersymmetry is diminished. At the boundary of the region allowed by
E821, the lightest detectable supersymmetric particle is the lighter stau
${\tilde \tau}_1$.  For our default choices $m_b(m_b)^{\overline
{MS}}_{SM} = 4.25$, $m_t = 175$~GeV and $A_0 = 0$ as in
Fig.~\ref{fig:upper}, and the test values $\tan \beta = 10, 30, 50, 55$,
we find that $m_{{\tilde \tau}_1} \la 190, 320, 420, 580$~GeV. We
therefore conclude that a first-generation linear $e^+ e^-$ collider with
centre-of-mass energy above 1.2~TeV would be `guaranteed' to find
supersymmetry within our CMSSM framework. A machine with centre-of-mass
energy above 800~GeV would be similarly `guaranteed' to find supersymmetry
if $\tan \beta \la 45$.

\begin{figure}
\vspace*{-0.75in}
\begin{minipage}{8in}
\epsfig{file=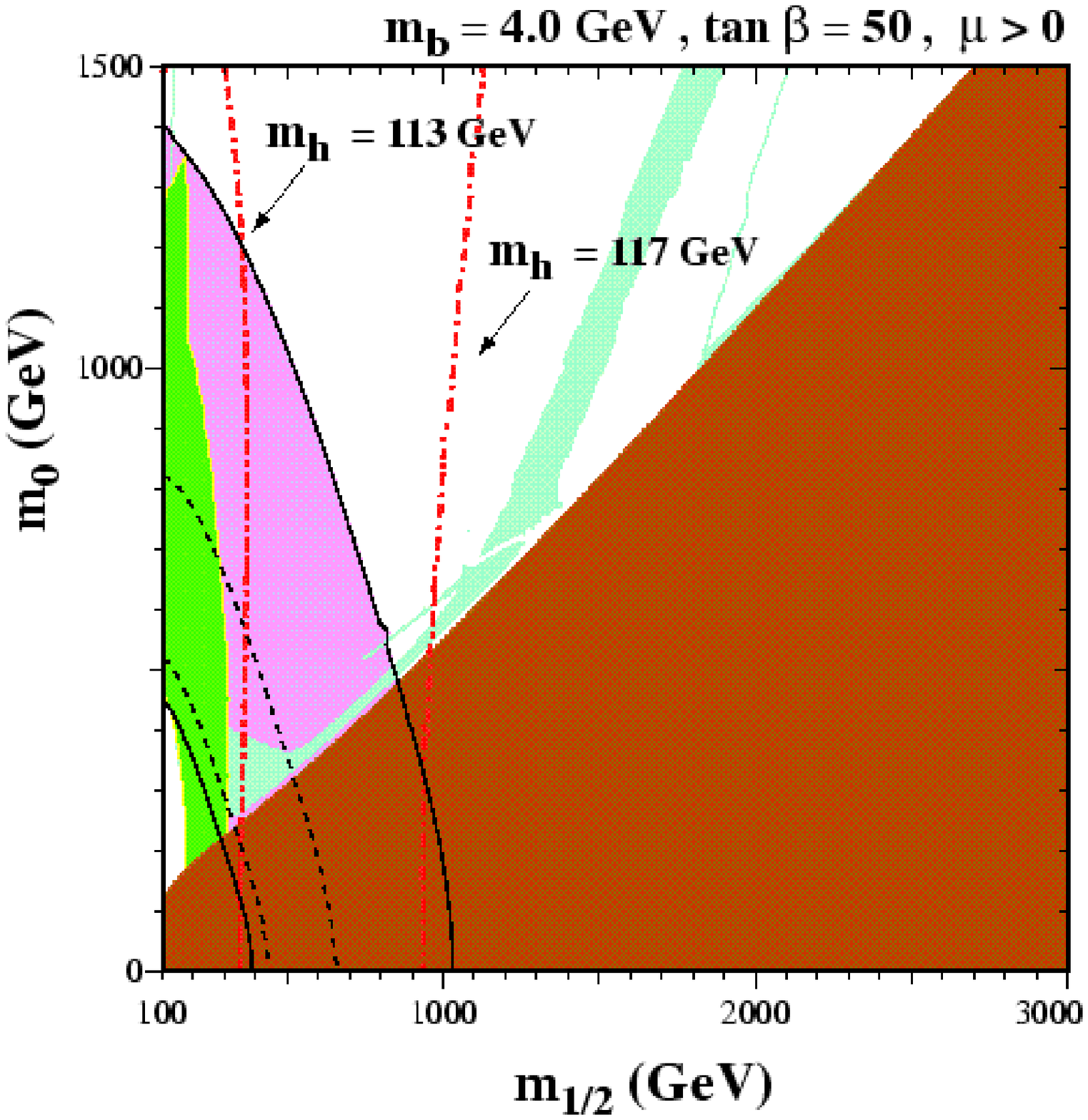,height=3.5in}
\epsfig{file=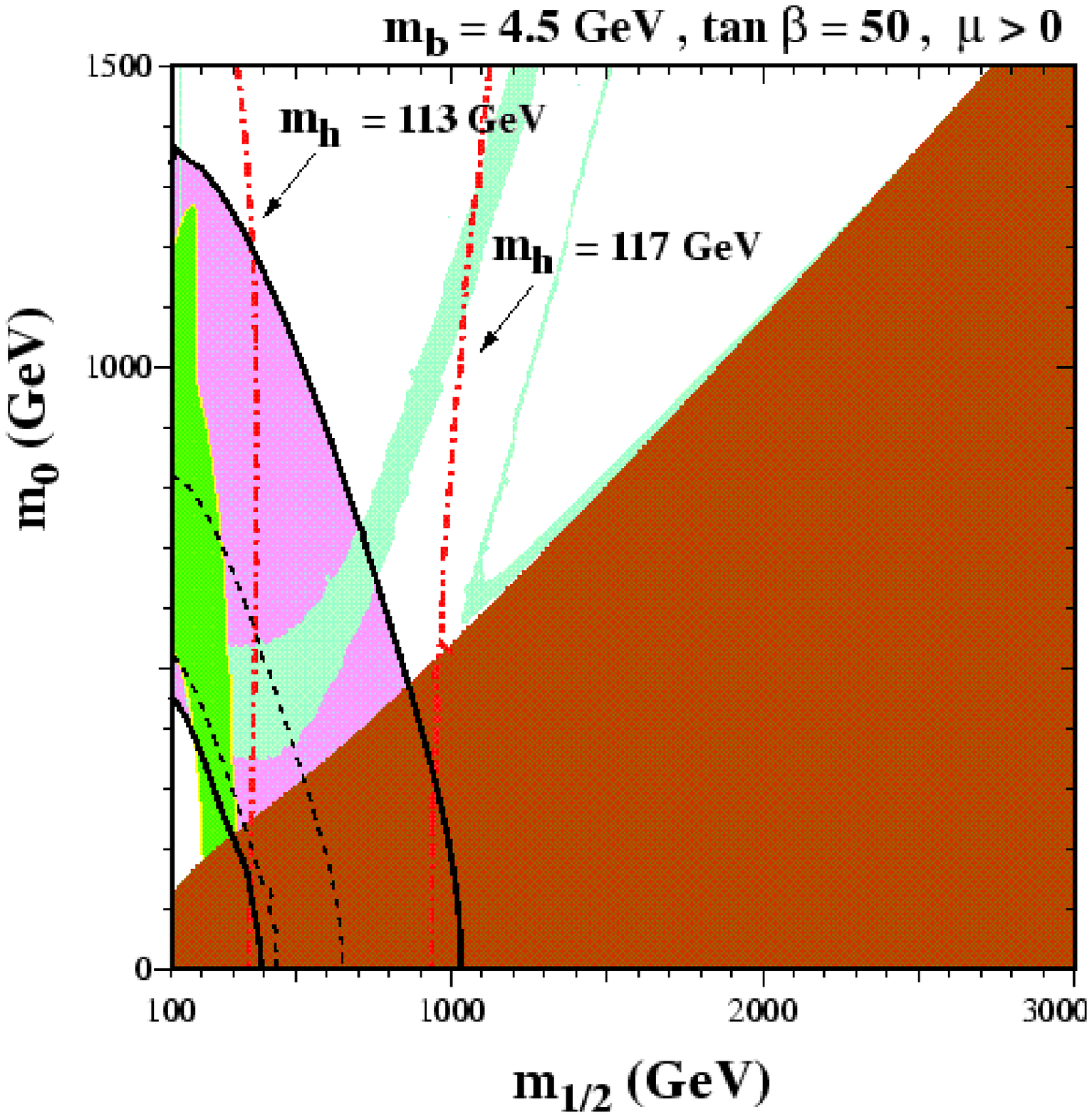,height=3.5in} \hfill
\end{minipage}
\begin{minipage}{8in}
\epsfig{file=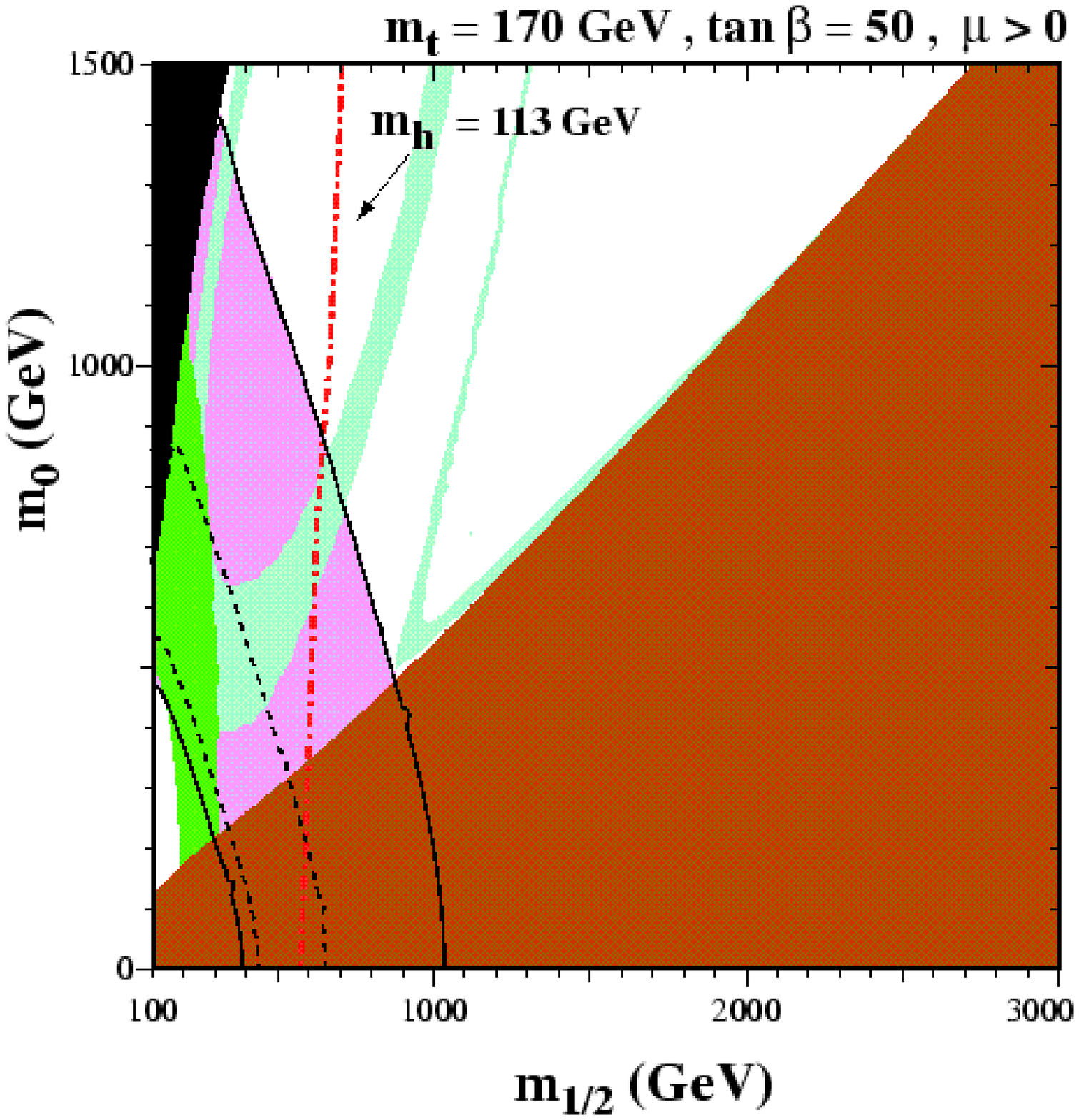,height=3.5in}
\epsfig{file=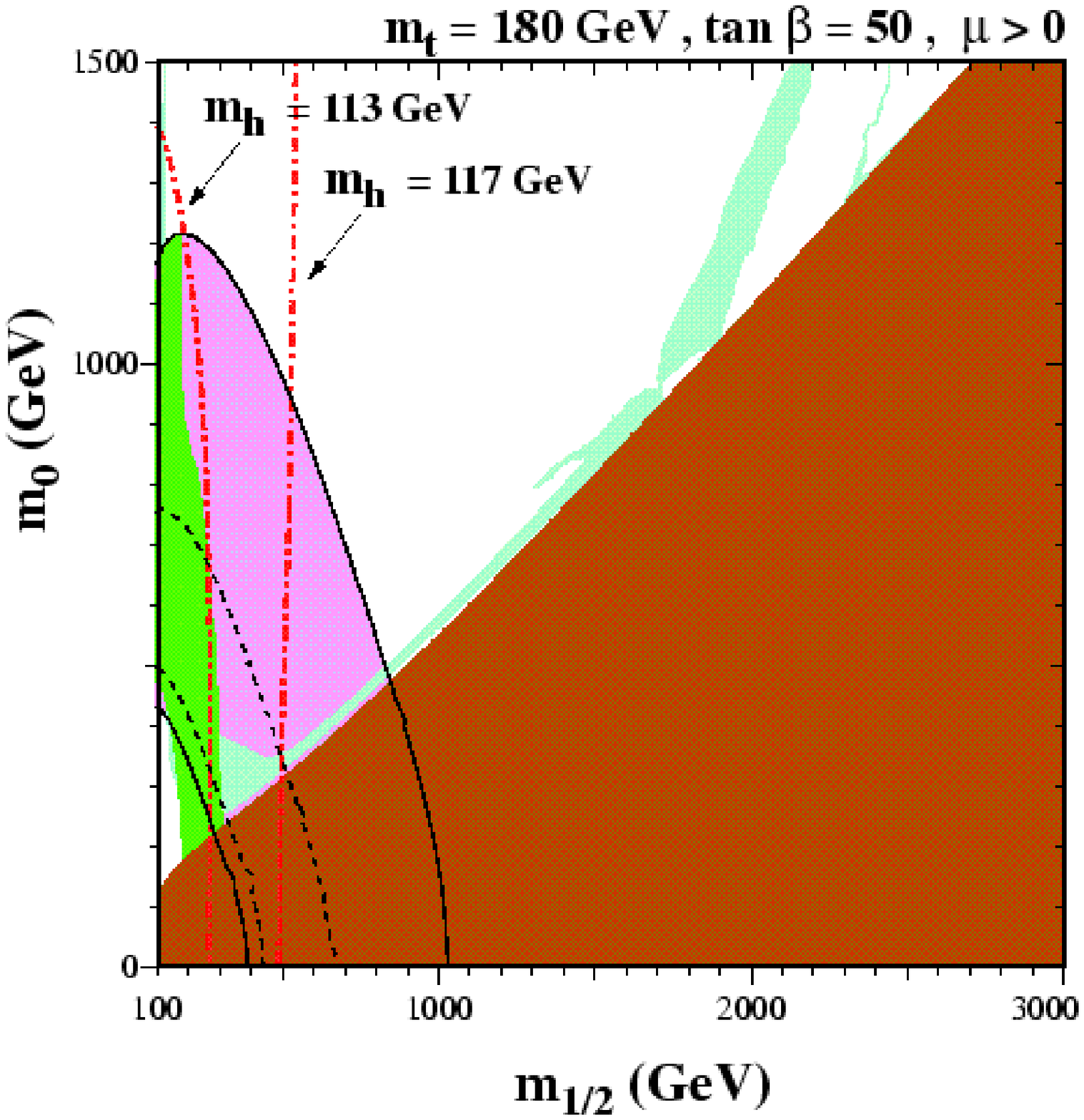,height=3.5in} \hfill
\end{minipage}
\caption{\label{fig:variations}
{\it Comparison between the $(m_{1/2}, m_0)$ planes for $\tan \beta = 50,
\mu > 0$ and $A_0 = 0$, with different values of other input parameters.
Panels (a) and (b) are for $m_t = 175$~GeV, $m_b(m_b)^{\overline
{MS}}_{SM} 
= 4.0$ and $4.5$~GeV, respectively. Panels (c) and (d) are for 
$m_b(m_b)^{\overline {MS}}_{SM} = 4.25$~GeV and $m_t = 170$ and $180$~GeV,
respectively.}}
\end{figure}

Fig.~\ref{fig:variations} for $\tan \beta = 50, \mu > 0$ and $A_0 = 0$
illustrates the effects of varying $m_b(m_b)^{\overline {MS}}_{SM}$
between (a) 4.0 and (b) 4.5~GeV, and of varying $m_t$ between (c) 170 and
(d) 180~GeV. As one would expect, the changes in the $\delta a_\mu$
constraint are minor, being essentially associated with changes in the RGE
and vacuum analysis. Also as expected, the $m_h$ contours and the $b \to s
\gamma$ constraint are rather similar in panels (a) and (b): the important
change as one varies $m_b$ is in the cosmological constraint. In
particular, the rapid $\chi \chi \to A, H$ annihilation `funnel' moves to
lower $m_{1/2}$ as $m_b$ increases, reducing the combined upper limit on
$m_{1/2}$ when $m_b(m_b)^{\overline {MS}}_{SM} = 4.5$~GeV, and increasing
the upper limit on $m_0$.  However, our overall conclusions on the
observability of the CMSSM at different colliders are unchanged. As seen
in panels (c) and (d), the main effects of varying $m_t$ are to move the
$m_h$ contours and the allowed cosmological region~\footnote{Note also the
black region in panel (c) of Fig.~\ref{fig:variations}, which is where we
find no consistent electroweak vacuum. There are similar but smaller
regions for larger $m_t$, that are not shown. The size of this forbidden
region is quite sensitive to the treatment of $m_t$, a topic we leave for
another occasion.}. As a result, the lower bound on $\tan \beta$ is
relaxed for $m_t = 180$~GeV. However, the effects on the bounds on
$m_{1/2}$ and $m_0$ in Fig.~\ref{fig:upper} are again relatively minor. We
do not display the effects of varying $-2 \times m_{1/2} \le A_0 \le 2
\times m_{1/2}$: the main changes are in the allowed cosmological region,
whose sensitivity to input assumptions were commented on
previously~\cite{EFGOSi}, but the effects on the bounds on $m_{1/2}$ and
$m_0$ in Fig.~\ref{fig:upper} are again not very important, though the
lower bound on $\tan \beta$ may again be relaxed.

In conclusion: we have combined the E821 $\delta a_\mu$ constraint with
other constraints on the CMSSM, including the LEP `signal' for the Higgs
boson, $b \to s \gamma$ and the favoured range of the cosmological relic
density $\ohsq$. We find a high degree of consistency for $\tan \beta \ga
10$, and interesting upper bounds on $m_{1/2}$ and $m_0$. There is a
corner of parameter space where the Fermilab Tevatron collider may find
supersymmetry. On the other hand, discovery of supersymmetry is
`guaranteed' at the LHC, within our stated theoretical assumptions.  The
E821 $\delta a_\mu$ constraint increases the chance that a
first-generation $e^+ e^-$ linear collider will find supersymmetry, though
there is no `guarantee' unless its centre-of-mass energy exceeds 1.2~GeV. 

{\it The E821 measurement of $\delta a_\mu$ provides an important
constraint on the CMSSM, and may already be the most promising positive
evidence for it. With the prospects of a significant reduction in the E821
error bar in the near future, we may be living in exciting times for
supersymmetry.}

{~}\\
\noindent
Finally, for completeness, we comment on recent papers related to
ours.

Ref.~\cite{EKRW} analyzes the BNL measurement but not the other
constraints discussed here. It is suggested that sparticles may be
produced at the Tevatron
in Run II, but their observability is not discussed. In this regard, as
discussed above, we are not very encouraged by previous
studies~\cite{Run2sugra}.

The most complete previous supersymmetric interpretation of the BNL
measurement is given in~\cite{FM}, which also includes some discussions of
the LEP Higgs, $b \to s \gamma$ and cosmological constraints. The main
difference between that work and ours is in the cosmology: the regions of
CMSSM parameter space given in~\cite{FM} do not include all the
coannihilation region, that extends in our calculations~\cite{EFGOSi} up
to $m_{1/2} = 1400 (2200)$~GeV for $0.1 \le \ohsq \le 0.3$ for $\tan \beta
= 10 (50)$. Conversely, the `focus point' region, that is now disallowed
by the BNL and other constraints, is beyond the domains of the $(m_{1/2},
m_0)$ plane that we plot. Our analysis for $\tan \beta = 50$ also
differs in the treatment of the direct-channel $A, H$ poles, that are
responsible for a large allowed region for $\ohsq$ in the analysis
of~\cite{FM}, but lead to narrow funnels in our analysis~\cite{EFGOSi}. 

The prospects for dark matter detection in the light of the BNL
measurement are discussed in~\cite{BG}, where LEP, $b \to s \gamma$ and
$\ohsq$ are also taken into account. The current LEP constraint $m_h \ge
113.5$~GeV was not used. This applies in the generic MSSM for $\tan \beta
\la 10$ and in the CMSSM at essentially all $\tan \beta$, excludes a large
range of $m_{1/2}$ and requires $m_\chi \ga 150$~GeV~\cite{EFGOSi}. We are
not in a position to compare treatments of $b \to s \gamma$.  The analysis
of $\ohsq$ in~\cite{BG} is based on DarkSUSY~\cite{DARKSUSY}, which does
not include all the effects at large $\tan \beta$ that we discussed
in~\cite{EFGOSi} and here. 

We use the same formulae~\cite{IN} as~\cite{CNnew} to implement the BNL
constraints on $m_0$ and $m_{1/2}$, with which we agree quite closely. 
Ref.~\cite{CNnew} also discusses the LEP Higgs `signal' and makes
qualitative comments on supersymmetric dark matter, but does not discuss
$b \to s \gamma$.

Ref.~\cite{KMY} discusses implications for the unconstrained MSSM and for
gauge-mediated models of supersymmetry breaking.

Ref.~\cite{HT} discusses the supersymmetric interpretation of $\delta
a_\mu$ in relation to models of neutrino masses and the observability of
$\mu \to e \gamma$ decay.

We note that anomaly-mediated supersymmetry breaking is considered
in~\cite{FM,CNnew}. 
We also note that non-supersymmetric interpretations of the E821 result
are discussed in~\cite{newothers}. 

\vskip 0.5in
\vbox{
\noindent{ {\bf Acknowledgments} } \\
\noindent  
We thank Geri Ganis for useful information, and Pran Nath for discussions. 
The work of D.V.N. was partially supported by DOE grant
DE-F-G03-95-ER-40917, and that of K.A.O. by DOE grant
DE--FG02--94ER--40823.}


\begin{thebibliography}{99}

\bibitem{BNL}
H.~N.~Brown {\it et al.}, Muon $g-2$ Collaboration,
hep-ex/0102017.

\bibitem{CM}
A.~Czarnecki and W.~J.~Marciano,
hep-ph/0102122; see also
hep-ph/0010194.

\bibitem{Davieretal}
M.~Davier and A.~H\"ocker,
Phys.\ Lett.\ B {\bf 435} (1998) 427.

\bibitem{Novo}
R.~R.~Akhmetshin {\it et al.}, CMD-2 Collaboration,
Phys.\ Lett.\ B {\bf 475} (2000) 190.

\bibitem{BES}
J.~Z.~Bai {\it et al.}, BES Collaboration,
hep-ex/0102003, and references therein.

\bibitem{CLEO}
S.~Anderson {\it et al.}, CLEO Collaboration,
Phys.\ Rev.\ D {\bf 61} (2000) 112002.

\bibitem{Davierpc}
M. Davier, S. Eidelman and A.~H\"ocker, private communication; see,
however,
F.~J.~Yndurain, hep-ph/0102312.

\bibitem{LEPHiggs}
ALEPH collaboration, R.~Barate {\it et al.}, Phys.\ Lett.\ {\bf B495}
(2000) 1 [hep-ex/0011045];\\
L3 collaboration, M.~Acciarri {\it et al.}, Phys.\ Lett.\ {\bf B495}
(2000) 18 [hep-ex/0011043];\\
DELPHI collaboration, P. Abreu {\it et al.}, 
Phys.\ Lett.\ B {\bf 499} (2001) 23;\\
OPAL collaboration, G. Abbiendi {\it et al.}, Phys.\ Lett.\ {\bf B499}
38.\\
For a preliminary compilation of the LEP data presented on Nov. 3rd, 2000,
see:\\
P. Igo-Kemenes, for the LEP Higgs working group,\\
{\tt http://lephiggs.web.cern.ch/LEPHIGGS/talks/index.html}.\\
For a recent compilation of other LEP search data, as presented on Sept.
5th, 2000, see:\\
T. Junk, hep-ex/0101015.

\bibitem{Fayet}
P. Fayet, {\it Unification of the Fundamental Particle Interactions}, eds.
S.~Ferrara, J.~Ellis and P.~van~Nieuwenhuizen (Plenum, New York, 1980),
p.587.

\bibitem{GM}
J.~A.~Grifols and A.~Mendez,
Phys.\ Rev.\ {\bf D26} (1982) 1809.


\bibitem{EHN}
J.~Ellis, J.~Hagelin and D.~V.~Nanopoulos,
Phys.\ Lett.\ {\bf B116} (1982) 283.

\bibitem{BM}
R.~Barbieri and L.~Maiani,
Phys.\ Lett.\ {\bf B117} (1982) 203.

\bibitem{Kosower}
D.~A.~Kosower, L.~M.~Krauss and N.~Sakai,
Phys.\ Lett.\ {\bf B133} (1983) 305.

\bibitem{AN}
T.~C.~Yuan, R.~Arnowitt, A.~H.~Chamseddine and P.~Nath,
Z.\ Phys.\ {\bf C26} (1984) 407.

\bibitem{Ven}
I.~Vendramin,
Nuovo Cim.\ {\bf A101} (1989) 731.

\bibitem{Abel}
S.~A.~Abel, W.~N.~Cottingham and I.~B.~Whittingham,
Phys.\ Lett.\ {\bf B259} (1991) 307.

\bibitem{LNW}
J.~L.~Lopez, D.~V.~Nanopoulos and X.~Wang,
Phys.\ Rev.\ {\bf D49} (1994) 366.

\bibitem{IN}
T.~Ibrahim and P.~Nath,
Phys.\ Rev.\ {\bf D62} (2000) 015004.

\bibitem{Moroi}
T.~Moroi,
Phys.\ Rev.\ {\bf D53} (1996) 6565;
M.~Carena, G.~F.~Giudice and C.~E.~Wagner,
Phys.\ Lett.\ {\bf B390} (1997) 234;
K.~T.~Mahanthappa and S.~Oh,
Phys.\ Rev.\ {\bf D62} (2000) 015012;
T.~Blazek,
hep-ph/9912460;
U.~Chattopadhyay, D.~K.~Ghosh and S.~Roy,
Phys.\ Rev.\ D {\bf 62} (2000) 115001;

\bibitem{CM2l}
A.~Czarnecki, B.~Krause and W.~J.~Marciano,
Phys.\ Rev.\ {\bf D52} (1995) 2619, and
Phys.\ Rev.\ Lett.\ {\bf 76} (1996) 3267.

\bibitem{CN}
U.~Chattopadhyay and P.~Nath,
Phys.\ Rev.\ {\bf D53} (1996) 1648.

\bibitem{bsgexpt}
CLEO Collaboration,
M.S. Alam et al., Phys.\ Rev.\ Lett.\ {\bf 74} (1995) 2885 as updated in
S.~Ahmed et al., {CLEO CONF 99-10};
BELLE Collaboration, BELLE-CONF-0003, contribution to the 30th
International conference on High-Energy Physics, Osaka, 2000.

\bibitem{EHNOS}
J. Ellis, J.S. Hagelin, D.V. Nanopoulos, K.A. Olive
and M. Srednicki, Nucl. Phys. {\bf B238} (1984) 453; see also
H. Goldberg, Phys. Rev. Lett. {\bf 50} (1983) 1419.

\bibitem{EFOS}
J.~Ellis, T.~Falk, K.~A.~Olive and M.~Schmitt,
Phys.\ Lett.\ {\bf B388} (1996) 97 and
Phys.\ Lett.\ {\bf B413} (1997) 355;
J.~Ellis, T.~Falk, G.~Ganis, K.~A.~Olive and M.~Schmitt,
Phys.\ Rev.\ {\bf D58} (1998) 095002.

\bibitem{EFOSi}
J.~Ellis, T.~Falk and K.~A.~Olive, Phys.\ Lett.\ {\bf
B444}, 367 (1998);  J.~Ellis, T.~Falk, K.~A.~Olive and M.~Srednicki,
Astropart.\ Phys.\ {\bf 13} (2000) 181. 

\bibitem{EFGO}
J.~Ellis, T.~Falk, G.~Ganis and K.~A.~Olive,
Phys.\ Rev.\ {\bf D62} (2000) 075010.

\bibitem{EGNO}
J.~Ellis, G.~Ganis, D.~V.~Nanopoulos and K.~A.~Olive,
hep-ph/0009355.

\bibitem{EFGOSi}
J.~Ellis, T.~Falk, G.~Ganis, K.~A.~Olive and M.~Srednicki,
hep-ph/0102098.

\bibitem{LNY}
J.~L.~Lopez, D.~V.~Nanopoulos and K.~Yuan,
Nucl.\ Phys.\ B {\bf 370} (1992) 445;
S.~Kelley, J.~L.~Lopez, D.~V.~Nanopoulos, H.~Pois and K.~Yuan,
Phys.\ Rev.\ {\bf D47} (1993) 2461;
J. Lopez, D.~V.~Nanopoulos and K.-J.~Yuan,
Phys.\ Rev.\ {\bf D48} (1993) 2766.

\bibitem{others}
A.~Bottino, V.~de Alfaro, N.~Fornengo, G.~Mignola and S.~Scopel,
Astropart.\ Phys.\ {\bf 1}, 61 (1992);
P.~Nath and R.~Arnowitt,
Phys.\ Rev.\ Lett.\ {\bf 70}, 3696 (1993);
G.~L.~Kane, C.~Kolda, L.~Roszkowski and J.~D.~Wells,
Phys.\ Rev.\ {\bf D49}, 6173 (1994);
R.~Arnowitt and P.~Nath,
Phys.\ Rev.\ {\bf D54}, 2374 (1996);
M.~Drees and M.~M.~Nojiri,
Phys.\ Rev.\ {\bf D47}, 376 (1993);
H.~Baer and M.~Brhlik,
Phys.\ Rev.\ {\bf D53} (1996) 597;
V.~Barger and C.~Kao,
Phys.\ Rev.\ {\bf D57} (1998) 3131;
J.~Edsjo and P.~Gondolo,
Phys.\ Rev.\ {\bf D56} (1997) 1879;
A.~B.~Lahanas, D.~V.~Nanopoulos and V.~C.~Spanos,
Phys.\ Lett.\  {\bf B464} (1999) 213 and
Phys.\ Rev.\ {\bf D62} (2000) 023515;
M.~E.~G\'omez,
G.~Lazarides and C.~Pallis,
Phys.\ Rev.\ {\bf D61}, 123512 (2000)
and
Phys.\ Lett.\ {\bf B487}, 313 (2000).

\bibitem{reviews}
For recent reviews, see: M.~Drees,
hep-ph/0101217;
R.~Arnowitt, B.~Dutta and Y.~Santoso,
hep-ph/0101020;
A.~Corsetti and P.~Nath,
hep-ph/0011313;
A.~Bottino, N.~Fornengo and S.~Scopel,
hep-ph/0012377;
L.~Bergstrom,
Rept.\ Prog.\ Phys.\ {\bf 63} (2000) 793;
G. Ganis,
hep-ex/0102013.

\bibitem{newbsgcalx}
C. Degrassi, P. Gambino and G.~F. Giudice, 
JHEP {\bf 0012} (2000) 009; see also
M.~Carena, D.~Garcia, U.~Nierste and C.~E.~Wagner,
hep-ph/0010003.

\bibitem{mb}
This is similar to the range quoted by
D.E. Groom {\it et al.}, Euro.\ Phys.\ J.\ {\bf C15} (2000) 1,
{\tt http://pdg.lbl.gov/}.

\bibitem{HHH}
For our numerical analysis, we use the results of
H.E. Haber, R. Hempfling and A.H. Hoang,
Zeit.\ f\"ur Phys.\ {\bf C75} (1997) 539; 
see also M. Carena, M. Quir\'os and C.E.M. Wagner, 
Nucl. Phys. {\bf B461} (1996) 407.
We have checked that these results agree, within the expected
uncertainties, with those of
M.~Carena, H.~E.~Haber, S.~Heinemeyer, W.~Hollik, C.~E.~Wagner and
G.~Weiglein,
Nucl.\ Phys.\ {\bf B580} (2000) 29.

\bibitem{Run2sugra}
S.~Abel {\it et al.}, SUGRA Working Group Collaboration,
hep-ph/0003154.

\bibitem{CMS}
S.~Abdullin and F.~Charles,
Nucl.\ Phys.\ {\bf B547} (1999) 60.

\bibitem{EGO}
J.~Ellis, G.~Ganis and K.~A.~Olive,
Phys.\ Lett.\ {\bf B474} (2000) 314.

\bibitem{Feng}
J.~L.~Feng, K.~T.~Matchev and F.~Wilczek,
Phys.\ Lett.\ {\bf B482} (2000) 388, and
Phys.\ Rev.\ {\bf D63} (2001) 045024.

\bibitem{EKRW}
L.~Everett, G.~L.~Kane, S.~Rigolin and L.~Wang,
hep-ph/0102145.

\bibitem{FM}
J.~L.~Feng and K.~T.~Matchev,
hep-ph/0102146.

\bibitem{BG}
E.~A.~Baltz and P.~Gondolo,
hep-ph/0102147.

\bibitem{DARKSUSY}
P.~Gondolo, J.~Edsjo, L.~Bergstrom, P.~Ullio and E.~A.~Baltz,
astro-ph/0012234.

\bibitem{CNnew}
U.~Chattopadhyay and P.~Nath,
hep-ph/0102157.

\bibitem{KMY}
S.~Komine, T.~Moroi and M.~Yamaguchi,
hep-ph/0102204.

\bibitem{HT}
J. Hisano and K. Tobe, hep-ph/0102315.

\bibitem{newothers}
K.~Lane,
hep-ph/0102131;
U.~Mahanta,
hep-ph/0102176;
D.~Chakraverty, D.~Choudhury and A.~Datta,
hep-ph/0102180;
T.~Huang, Z.~H.~Lin, L.~Y.~Shan and X.~Zhang,
hep-ph/0102193;
D.~Choudhury, B.~Mukhopadhyaya and S.~Rakshit,
hep-ph/0102199;
U.~Mahanta,
hep-ph/0102211;
S.~N.~Gninenko and N.~V.~Krasnikov,
hep-ph/0102222;
K.~Cheung,
hep-ph/0102238;
P.~Das, S.~K.~Rai and S.~Raychaudhuri,
hep-ph/0102242;
T.~W.~Kephart and H.~P\"as,
hep-ph/0102243.
E.~Ma and M.~Raidal,
hep-ph/0102255;
Z.~Xiong and J.~M.~Yang,
hep-ph/0102259;
A.~Dedes and H.~E.~Haber, hep-ph/0102297;
Z.~Z.~Xing, hep-ph/0102304;
H. Senju, Nagoya City University preprint NMWJ~28 (Feb. 2001).
\end{thebibliography}
\end{document}